\documentclass[aps,prl,twocolumn,floatfix,showpacs]{revtex4}
\usepackage{graphicx}
\usepackage{verbatim}
\usepackage{ulem}
\usepackage{color}

\begin{document}

\title{ Polarization-driven topological insulator transition in a
GaN/InN/GaN quantum well}
\author{M. S. Miao$^{1,3}$}
\email{miaoms@engineering.ucsb.edu}
\author{Q. Yan$^{1}$}
\author{C. G. Van de Walle$^{1}$}
\author{W. K. Lou$^{2}$}
\author{L. L. Li$^{2}$}
\author{K. Chang$^{2,3}$}
\email{kchang@semi.ac.cn}
\affiliation{$^{1}$Materials Research Laboratory and Materials Department, University of
California, Santa Barbara, CA 93106-5050}
\affiliation{$^{2}$SKLSM, Institute of Semiconductors, Chinese Academy of Sciences, P. O.
Box 912, Beijing 100083, China}
\affiliation{$^{3}$Beijing Computational Science Research Center, Beijing 10084, P. R.
China}


\begin{abstract}
Topological insulator (TI) states have been demonstrated in materials with narrow gap and large spin-orbit interactions (SOI).
Here we demonstrate that nanoscale engineering can also give rise to a TI state,
even in conventional semiconductors with sizable gap and small SOI.
Based on advanced first-principles calculations combined with an effective low-energy \textbf{k$ \cdot $p}
Hamiltonian, we show that the intrinsic
polarization of materials can be utilized to simultaneously reduce the energy gap and
enhance the SOI, driving the system to a TI state. The proposed system consists of ultrathin InN layers embedded into GaN,
a layer structure that is experimentally achievable.
\end{abstract}

\maketitle

\begin{figure}[h!]
\includegraphics[width=8.5cm]{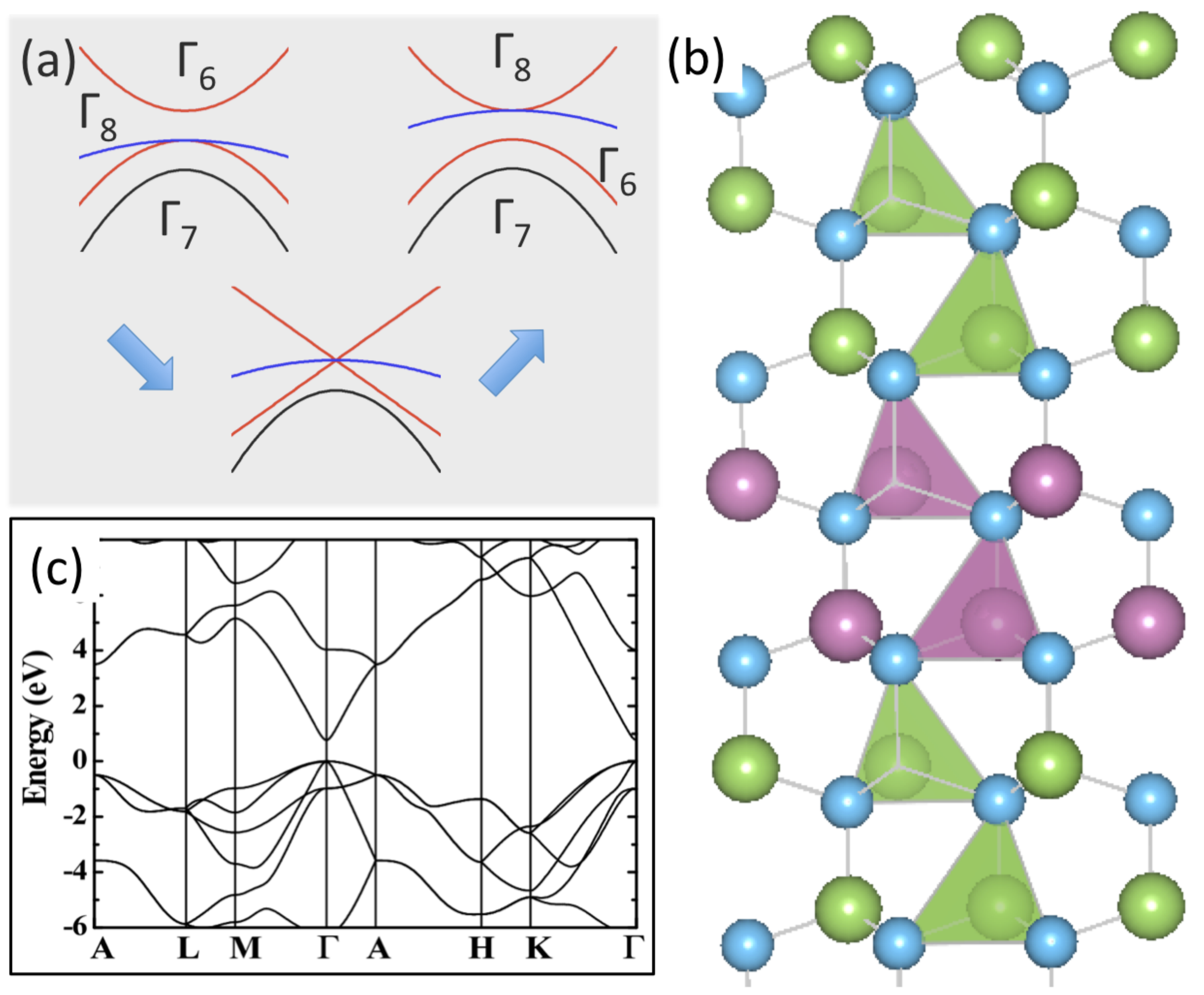}
\caption{{\protect\footnotesize (color online) (a) Counterclockwise, bands
of normal semiconductors such as CdTe, bands at the zero-gap transition
point, and inverted bands of semiconductors. (b) Side view of a GaN/InN/GaN QW
including two atomic layers of InN in a GaN matrix along the (0001) polar
orientation. The red, green and blue balls represent the In, Ga and N atoms. (c) Calculated band structure of
bulk InN using the HSE hybrid functional. }}
\label{2DInN}
\end{figure}

Topological insulators (TIs), a new state of quantum matter,
have recently attracted significant attention, both for their fundamental research interest and for their
potential device applications \cite{RMP,Qi2011,Kane2005}. The TI state can be achieved in
semiconductors with inverted bands [Fig.~\ref{2DInN}(a)] and is usually
driven by the intrinsic spin-orbit interaction (SOI) arising from heavy host
atoms \cite{BHZ,Konig2007}. In the TI state the system features an
insulating bulk and gapless surface/edge states. These novel states show a
linear Dirac spectrum that is protected by time-reversal symmetry,
resulting in quantized and robust conductance when the Fermi level is in the
gap.

This new quantum state of electrons was first proposed theoretically
for quantum wells (QWs) and then for three-dimensional cases, and was later
demonstrated experimentally by angle-resolved photoemission spectroscopy
(ARPES) \cite{3DTI,Hsieh2008}.  Early efforts to find TI materials
focused on binary II-VI compounds or chalcogenides with heavy atoms, such as
HgTe and Bi$_{2}$X$_{3}$ (X=Te, Se) \cite{Hsieh2008}. By adjusting the thickness or strain of a
CdTe/HgTe/CdTe QW, or applying an external electric field, one can drive the
system into the TI phase. The investigations have been extended to ternary Heusler compounds,
with first-principles calculations playing an important role \cite{Chadov2010,Lin2010,Franz2010}.

In spite of such progress, the search for TIs in more commonly used
semiconductor systems has remained elusive, holding out the prospect of integration with
conventional semiconductor devices. However, commonly used
semiconductors typically have sizable band gaps and insignificant SOI, in
contrast to the requirements for a TI phase transition. One promising approach
is to apply a strong gate voltage in a two-dimensional nano-scale
QW system \cite{Yang2008,Li2009}.  The strong electric field
reduces the gap and induces considerable Rashba SOI
that may transform the system into a TI. The major difficulty of
this approach is that the required electric field is very strong and
typically reduces the gap by less than 0.1 eV.

\begin{figure}[tbp]
\includegraphics[width=9cm]{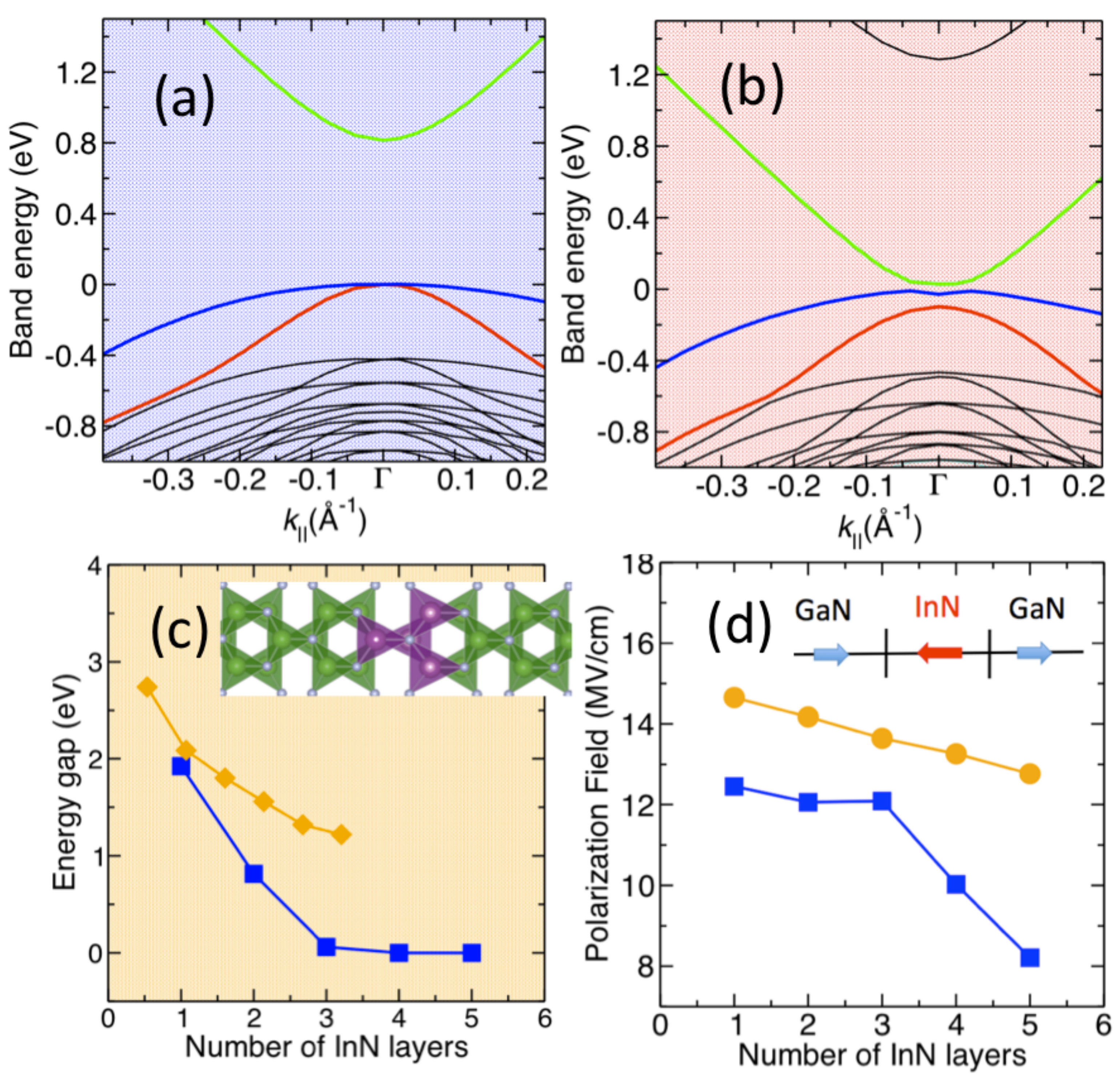}
\caption{{\protect\footnotesize (color online)  (a) and (b) Band structure of a
GaN/InN/GaN QW around the $\Gamma $ point for 2 and 4 ML of InN,
based on first-principles DFT-HSE calculations. The green lines represent electron states,
red lines light-hole states, and blue lines heavy-hole states.
(c) Calculated energy gaps as a function of number
of InN layers for polar ([0001]) (blue squares) and nonpolar
($[10 \bar{1}0]$) (orange circles) GaN/InN/GaN QWs. The inset shows a $[10\bar{1}0]$ QW
with two InN layers.
(d) Polarization field as a function of number of inserted
InN layers calculated by DFT-HSE (blue squares) and based on theoretical polarization constants
(orange circles). The blue and red block arrows in the inset show the polarization direction of GaN and InN regions \cite{suppl}.} }
\label{bands}
\end{figure}

Here we propose that the intrinsic polarization of materials can be
utilized to reduce the energy gap and invert the band ordering at the $\Gamma$ point.
The wurtzite structure of the III-nitrides allows for the presence of spontaneous and
piezoelectric polarization fields in structures grown along the [0001] direction.
The piezoelectric fields arise due to the large strain that is induced in a thin InN
layer grown pseudomorphically on GaN (i.e., maintaining the in-plane lattice constant of the GaN layer;
we assume growth on a GaN substrate or thick GaN layer).
We show by first-principles electronic structure calculations that
this polarization can invert the bands of a GaN/InN/GaN QW when the InN
thickness exceeds three monolayers (ML)--a ML being defined as a double layer
of In and N (with a thickness of 3.12 {\AA} in strained InN) along the [0001] axis.
We then use an eight-band
\textbf{k$\cdot $p} model that includes both spin-orbit interaction (SOI) and
the strong polarization field to prove that such a system can
become a TI and possess edge states in the energy gap of a Hall bar
structure.

The proposed TI has many unique advantages, including: \
1) it can be realized based on commonly used semiconductors and
be integrated into various devices; 2) it is driven by
large intrinsic polarization fields; 3) the TI state can be
manipulated by applying external fields or injecting charge carriers
and can be adjusted by standard semiconductor techniques, including
doping, alloying and varying the QW thickness; 4) the
polarization field can induce a large Rashba SOI in this system
containing only light elements, which provides a new
approach to manipulating spin freedom in such systems.

Our proposed QW consists of InN layers sandwiched between GaN along
the [0001] direction [Fig.~\ref{2DInN}(b)]. Group-III nitrides (III-N),
including InN, GaN and AlN, have been intensively studied
because of their applications in light emitters, high-frequency transistors,
and many other areas. An important feature of
III-N compounds is that although their band gap varies from
0.7 eV to 6.2 eV, all three compounds and their alloys are stable in
the same wurtzite structure with relatively
modest variations in lattice constant.  This feature, combined with advanced
growth techniques, allows the fabrication of high-quality
alloys and heterojunctions. InN has a fundamental gap of only $\sim$0.7 eV (Ref.~\onlinecite{Wu2002}), resulting in
strong coupling between the electron and hole states and
leading to a large non-parabolicity of the bands around the $\Gamma$ point
[Fig.~\ref{2DInN}(c)]. Correspondingly, its electron effective mass
(relevant for transport within the plane of the QW) has the exceedingly small value of
0.067$m_{0}$ (Ref.~\onlinecite{Rinke2008}). InN therefore has
strong potential for applications requiring highly mobile carriers, such as
terahertz devices or high-electron mobility transistors.
Growth of high-quality InN layers within a GaN matrix has been reported \cite{Yoshikawa2007},
and light emission from electron-hole recombination in such structures was observed \cite{Che2009}.

Our first-principles calculations are based on density functional theory (DFT)
with a plane-wave basis set and
projector augmented waves \cite{PAW}, as implemented in the VASP program \cite{VASP}.
Because the accuracy of the band gap is of key importance for this work,
we employ a hybrid functional \cite{HSE}.
Recent calculations have demonstrated the reliability of this approach for
producing structural parameters as well as band gaps in agreement with
experiment. Using a standard mixing parameter of
0.25, we found the band gaps of GaN and InN to be 3.25 and 0.62 eV (Ref.~\onlinecite{Moses2010}).
The lattice parameters $a$=$b$ and $c$ are found to be
3.182 \AA ~ and 5.175 \AA ~ for GaN and 3.548 \AA ~ and 5.751 \AA ~ for InN.
The GaN/InN/GaN QW is modeled by a supercell consisting of 1 to 5 atomic layers of InN and 23 or 24 atomic layers of GaN
(periodicity requires that the total number of atomic layers be even).
The SOI is not included in the first-principles calculations, but
its effect is included in the \textbf{k$\cdot $p} model.

Figures~\ref{bands}(a) and (b) show band structures around the $\Gamma$
point calculated with DFT using the hybrid functional of
Heyd, Scuseria and Ernzerhof (HSE) \cite{HSE}, and Fig.~\ref{bands}(c) shows the
band gap as a function of number of InN layers.
At 2 ML, the system has a gap of 0.82 eV (larger than the fundamental gap
due to strain and quantum confinement), but at 3 ML this is reduced to 0.06 eV. In both these cases, the band
structures are still normal in the sense that the heavy hole (HH) and light
hole (LH) states are degenerate at the $\Gamma$ point and are lower in energy
than the electron state (E1). At 4 ML, however, the system exhibits an inverted band structure, in which
 a gap of 0.10 eV is opened between
the HH and the LH states. Such an inverted band structure is a signature
of the transition to a TI state.

We now discuss the details of this transition to an inverted band structure.
For an ultrathin QW, the gap between the
valence and conduction states is determined by the interplay of three
factors, namely quantum confinement, polarization field, and
strain. The quantum-confinement effect is large for these thin QWs,
explaining why the 1ML and 2ML QWs have gaps
larger than bulk InN. However, the effect on the band gap becomes smaller than 0.1 eV when
the thickness of the InN QW exceeds 3 ML.

The effect of polarization, on the other hand, increases with increasing QW
thickness.  The difference in spontaneous polarization between InN and GaN, along with
the piezoelectric polarization induced by the strain in InN, generates polarization charges
[Fig.~\ref{bands}(d)] that give rise to an electric field in the InN layer.
(Depending on the details of the structure and the boundary conditions,
electric fields may also be present in the surrounding GaN layers.)
This polarization field leads to a potential drop over the InN QW, and
localizes hole states and electron states on opposite sides of the QW.
As the QW width increases the electron and hole states become closer in energy,
leading to a significant reduction in overall band gap [Fig.~\ref{bands}(c)].

In order to demonstrate that the band inversion is driven by
the polarization fields we also performed calculations for GaN/InN/GaN QWs in a nonpolar
($[10\bar{1}0]$) orientation. Quantum-confinement effects and
strains are similar for the two orientations, so the enhanced reduction in band gap for the
polar QW observed in Fig.~\ref{bands}(c) can be attributed to the presence of polarization fields.
In contrast, the gap for the nonpolar QW saturates to a value around 1 eV (different from the bulk gap due
to the presence of compressive strain in the InN layer).

The electric field in the InN layers [Fig. \ref{bands}(d)] is as large as 12.5 MV/cm.
This field is in principle due to strain-induced piezoelectric
polarization as well as spontaneous polarization.
Because the spontaneous polarization constants of GaN and InN are quite similar, and
because the strains are very large in these pseudomorphic layers,
the piezoelectric field dominates.

The first-principles results included in Fig.~\ref{bands}(d) show a decrease in the
electric field when the QW thickness increases from 1 to 3 ML (see Supplemental Material \cite{suppl} for a discussion
of critical layer thickness).
This is attributed to the finite thickness of the GaN in the calculations,
which really represent a GaN/InN superlattice.  The same effect is observed
in model calculations \cite{Fiorentini1999} [also shown in Fig.~\ref{bands}(d)]
based on theoretical spontaneous polarization and piezoelectric constants for
bulk GaN and InN \cite{Bernardini1997}.  The difference between the model
and the first-principles results for 1-3 ML is likely due to the fact that the strains
in InN are so large as to be outside the linear regime, linearity being an assumption in the
model calculations.

The additional decrease in field when the QW thickness exceeds 3 ML is due to
another effect not captured in the model, namely that when the
potential drop over the InN layer exceeds its band gap,
charge transfer occurs from valence-band states on one side of the well
to conduction-band states on the other, leading to screening of the
polarization field.  The onset of this effect coincides with the
point where the energy gap of the overall system goes to zero [Fig.~\ref{bands}(c)].

In order to further examine whether our proposed structure can undergo a topological
quantum phase transition, we constructed a six-band effective
model Hamiltonian based on the eight-band \textbf{k$\cdot $p} model (Kane model)
for wurtzite semiconductors. Distinct from the HgTe system \cite{BHZ}, 
however, the HH and LH states in the InN system are close in energy.
Therefore, a six-band effective low-energy Hamiltonian is used here,
expressed by the basis $\left \vert E,\uparrow \right \rangle $, $\left \vert
LH,\uparrow \right \rangle $, $\left \vert HH,\uparrow \right \rangle $,
$\left \vert E,\downarrow \right \rangle $, $\left \vert LH,\downarrow
\right \rangle $, $\left \vert HH,\downarrow \right \rangle $; its exact form is
described in the Supplemental Material \cite{suppl}.

The \textbf{k$\cdot$p} simulations show that the band structure of the GaN/InN/GaN QW is inverted
when the QW width is larger than 15.5 \AA. This thickness, corresponding to about 5 InN layers, is somewhat larger than
that of 3 to 4 InN MLs obtained in the first-principles calculations \cite{suppl}.
However, this difference does not
affect the key features of the TI transition.
The \textbf{k$\cdot $p} model enables us to discuss the interplay of
several important factors such as strain, polarization and spin-orbit
interaction and to reveal the underlying physics of the TI transition.
It allows us to to demonstrate the presence of
edge states in a spin Hall bar, which is a direct way of identifying the TI phase.
There are always an odd number of
Kramers pairs of edge states on the boundary of a topological insulator \cite{RMP,Qi2011}.
The edge-state energy spectrum can be obtained by solving the effective
six-band model for a Hall bar structure with a finite width.

\begin{figure}[tbp]
\includegraphics[width=9cm]{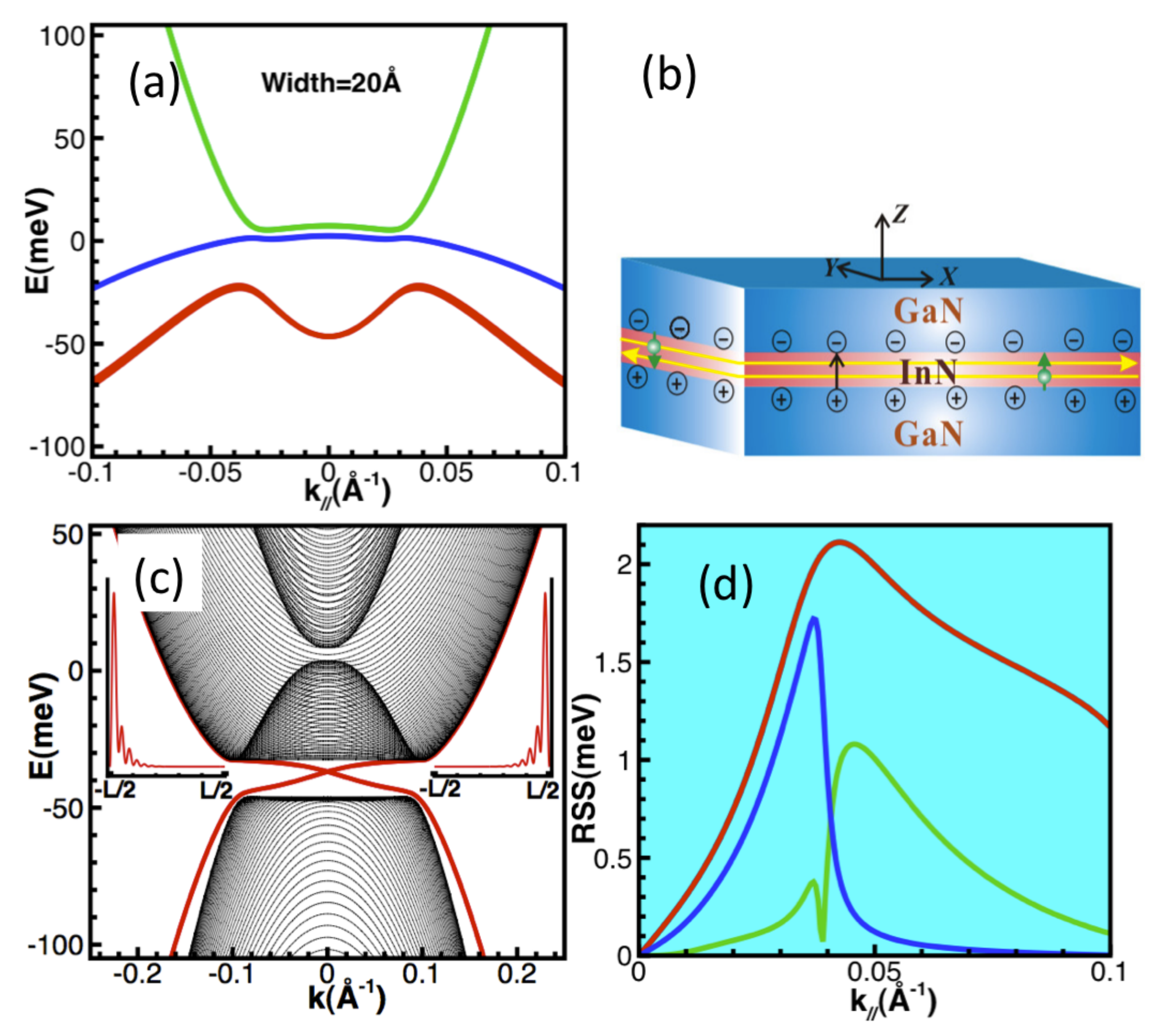}
\caption{{\protect\footnotesize (color online)
(a) Band structure of a 16 {\AA} GaN/InN/GaN QW obtained from the eight-band \textbf{k$\cdot $p} model.
(b) Schematic of a spin Hall bar with a width (along $y$) of 1000 \AA. The thickness along the [0001] growth direction (labeled as $z$)
comprises the InN QW plus two 200-{\AA}-thick GaN barrier layers on either side. The yellow closed circuit lines
show the helical edge states, and the green arrows show the
spin orientation. The short black arrows indicate the polarization-induced electric field.
(c) Band structure of the Hall bar obtained by
solving the effective six-band model. The insets show the density
distributions of one Kramers pair of
edge states: on the left the spin-up state at
$k_{\parallel}$=$-$0.1 {\AA}$^{ -1}$, on the right the spin-down state at
$k_{\parallel}$=0.1 {\AA}$^{-1}$.
(d) Rashba spin splitting of electron (green), HH (blue)
and LH (red) subbands.} }
\label{hall_bar}
\end{figure}

To illustrate the TI state for a GaN/InN/GaN QW, we calculated the edge states of a Hall bar
fabricated perpendicular to the growth plane [Fig.~\ref{hall_bar}(b)]. The resulting
band structure is presented in Fig.~\ref{hall_bar}(c) together with the density
distribution of a Kramers pair of edge states.
The presence of edge states in the gap between E1 and LH1 subbands is clearly demonstrated, and the
distribution of these states in real space shows that they are highly localized in the
vicinity of the edges of the Hall bar. These edge states are topologically
invariant under scattering, and therefore the corresponding mean free paths of
the carriers can be exceedingly large. We note that
although the HH band has only a slight effect on the formation of the TI state,
its inclusion is important when considering
the experimental detection of edge states. Like in other TI systems \cite{Du2011},
the HH band in a quantum spin Hall bar can create flat subbands and
conceal the topological edge states. However, the edge states can still be
observed in the minigap between the E1 and HH1
bands [see Fig. 3(a)]. This minigap is found to be 15 meV in our system.  While small, this is
larger than the hybridized minigap in the InAs/GaSb QW system in which
edge states have recently been observed \cite{Du2011}.

Spin-orbit interaction is essential in the transition to a TI state.
The intrinsic SOI in both InN and GaN is very small, on the order of
a few meV. However, in the vicinity of an inverted-band transition, a small SOI
is sufficient to drive the TI transition. Furthermore, we found that the
large polarization field induces a considerable Rashba SOI. The Rashba SOI in QWs has
attracted attention because it can be adjusted by gate
voltage \cite{Nitta1997,Grundler2000} and band engineering \cite{Winkler2003,Zawadzki2004}.
It thus provides a controllable approach to enhancing spin-polarized
transport in non-magnetic semiconductors. From the eight-band Kane model, we
estimate the strength of this Rashba SOI to be on the order of 1 to 2 meV [Fig. 3(d)].
This is comparable to the Rashba SOI induced by
an external electric field in InAs and HgTe quantum wells \cite{Yang2008,Li2009}.
A Rashba SOI of this magnitude usually occurs only in systems containing heavier
atoms. The unusually large Rashba SOI in GaN/InN/GaN QWs is due to the strength
of the polarization field, which easily exceeds 10 times the strength of
an applied electric field resulting from a gate voltage.

The proposed inverted band semiconductor structure based on thin InN
QWs and the presence of a TI state offer considerable advantages over
other systems including graphene \cite{Kane2005}, the Bi chalcogenides \cite{Hsieh2008} and the Heusler compounds \cite{Chadov2010}.
GaN/InN/GaN QW structures can be integrated in nitride-based transistors, which are
already extensively used in high-frequency and
high-power devices \cite{Mishra2008}. Because charge carriers
can screen the polarization fields, the polarization potential can
be controlled by adjusting the carriers densities in the QW, and therefore the
TI transition can be controlled by doping or applying a bias voltage.

In summary, based on first-principles
calculations and an effective low-energy \textbf{k$\cdot $p}
model, we have demonstrated that ultrathin GaN/InN/GaN QWs can undergo an inverted band
transition and become a topological insulator. This quantum phase
transition is driven by the strong polarization fields originating from the
wurtzite symmetry and lattice mismatch, and by the Rashba
spin-orbit interaction resulting from the field. This is the
first demonstration of the formation of a TI phase
caused by intrinsic polarization in commonly used semiconductors
with weak intrinsic SOI. Since polarization fields occur
in many materials, a similar mechanism may apply to other systems as well. Our
approach may pave the way toward integrating controllable TIs with
conventional semiconductor devices.

We thank Prof. X.-L. Qi for valuable discussions. MSM was supported as part of the Center for
Energy Efficient Materials, an Energy Frontier Research Center funded by the U.S. DOE-BES (DE-SC0001009).
QY was supported by the UCSB Solid State Lighting and Energy Center.
CVdW was supported by NSF (DMR-0906805).
LLL and KC were supported by NSFC No. 10934007 and 2011CB922204. The electronic
structure calculations made use of NSF-funded TeraGrid resources
(DMR07-0072N).

\end{document}